# Vector Magnetic Current Imaging of an 8 nm Process Node Chip and 3D Current Distributions Using the Quantum Diamond Microscope

Sean M. Oliver,[1,2,*] Dmitro J. Martynowych,[1,*] Matthew J. Turner,[2,3,4] , David A. Hopper,[1] Ronald L. Walsworth,[2,3,4] and Edlyn V. Levine[1,2,5,§]

[1]The MITRE Corporation, McLean, VA 22102, United States of America

[2]Quantum Technology Center, University of Maryland, College Park, Maryland 20742, United States of America

[3]Department of Electrical and Computer Engineering, University of Maryland, College Park, Maryland, 20742, United States of America

[4]Department of Physics, University of Maryland, College Park, Maryland, 20742, United States of America

[5]Department of Physics, Harvard University, Cambridge, MA 02138, United States of America

*These authors contributed equally to this work

§Corresponding author: evlevine@mitre-engenuity.org

## Abstract

The increasing trend for industry adoption of three-dimensional (3D) microelectronics packaging necessitates the development of new and innovative approaches to failure analysis. To that end, our team is developing a tool called the quantum diamond microscope (QDM) that leverages an ensemble of nitrogen-vacancy (NV) centers in diamond for simultaneous wide field-of-view, high spatial resolution, vector magnetic field imaging of microelectronics under ambient conditions [1,2]. Here, we present QDM measurements of two-dimensional (2D) current distributions in an 8 nm process node flip chip integrated circuit (IC) and 3D current distributions in a custom, multi-layer printed circuit board (PCB). Magnetic field emanations from the C4 bumps in the flip chip dominate the QDM measurements, but these prove to be useful for image registration and can be subtracted to resolve adjacent current traces on the micron scale in the die. Vias, an important component in 3D ICs, display only $B_x$ and $B_y$ magnetic fields due to their vertical orientation, which are challenging to detect with magnetometers that traditionally only measure the $B_z$ component of the magnetic field (orthogonal to the IC surface). Using the multi-layer PCB, we demonstrate that the QDM's ability to simultaneously measure $B_x$, $B_y$, and $B_z$ magnetic field components in 3D structures is advantageous for resolving magnetic fields from vias as current passes between layers. The height difference between two conducting layers is determined by the magnetic field images and agrees with the PCB design specifications. In our initial steps to provide further z depth information for current sources in complex 3D circuits using the QDM, we demonstrate that, due to the linear properties of Maxwell's equations, magnetic field images of individual layers can be subtracted from the magnetic field image of the total structure. This allows for isolation of signal from individual layers in the device that can be used to map embedded current paths via solution of the 2D magnetic inverse. Such an approach suggests an iterative analysis protocol that utilizes neural networks trained with images containing various classes of current sources, standoff distances, and noise integrated with prior information of ICs to subtract current sources layer by layer and provide z depth information. This initial study demonstrates the usefulness of the QDM for failure analysis and points to technical advances of this technique to come.

## Introduction

The continual development of microelectronic architectures and packaging are driving the industry beyond Moore's law. In order to increase performance and minimize footprint, microelectronics have gone into the third dimension by connecting multiple die vertically with through-silicon vias (TSVs) or Cu-Cu connections [3]. At the die scale, transistor utilization can be increased by routing power through the backside for beyond 5 nm node CMOS [4] and is the enabling technology behind Intel's recently announced PowerVia [5]. These advancements present a unique challenge for failure analysis as both methods increase the amount of metallization between the active gate layer and the backside silicon surface, thus preventing optical access to the gate layers. To ensure that failure analysis can detect defects, novel sensors that can function through multiple die and dense metal layers must be developed. Here, we present our progress towards this goal as we develop sensitive magnetic mapping with the quantum diamond microscope (QDM), which provides an image of the vector magnetic field produced by current traveling through integrated circuits (ICs) and printed circuit boards (PCBs).

The engineering complexity of 3D microelectronics packaging introduces many avenues for failure and isolating these faults poses a significant challenge as failures can originate from different die, assembly layers, or their interconnects [6]. Locating failures begins by determining whether there are shorts or opens where they are not expected. Localization in 3D space allows for subsequent high-resolution methods, such as scanning electron microscopy (SEM) and transmission electron microscopy (TEM) to determine the root cause of failure. Leading contenders for determining fault location are lock-in thermography (LIT) and optical beam induced resistance change (OBIRCH) [6]. Both methods rely on imaging infrared light in the IC to either induce localized heating (OBIRCH) or observe the thermal emission of a hotspot (LIT) [7]. Both techniques have their drawbacks. OBIRCH is a scanning modality and throughput time for high resolution scans can be cumbersome. LIT requires sensitive cameras and long integration times [8]. Both methods necessitate that the backside of the package is IR transparent, a requirement that is violated by multi-stack die and backside power delivery. In addition, these methods cannot detect true opens.

Magnetic field imaging has the potential to circumvent these issues. Current traveling through the IC or PCB produces a small vector magnetic field. Given a sufficiently sensitive magnetometer and adequate spatial resolution, the current path can be inferred from this information. Existing DC magnetic sensing technology utilizes superconducting quantum interference devices (SQUID) [9,10] and giant magnetoresistance (GMR) magnetometers [11]. SQUIDs are also capable of locating opens by detecting an AC signal propagating through the path [12–14]. However, these methods require point-by-point scanning to build up images, require cryogenics and vacuum shrouds in the case of SQUIDs, and can only detect the out-of-plane field component ($B_z$), which prevents current detection in TSVs or Cu-Cu interconnects. These factors combine to make imaging large footprint ICs challenging and reduces the effectiveness in 3D ICs with many active layers.

An image of a prototype QDM is shown in **Fig. 1(a)** and further detail is shown in the schematic of **Fig. 1(b)**. The QDM leverages nitrogen-vacancy (NV) center point defects in diamond, characterized by a substitutional nitrogen atom and a neighboring lattice vacancy (**Fig. 1(c)**) in the diamond carbon lattice. NV centers exhibit an optical response to varying magnetic fields that can be detected with visible imaging optics (see **Supplemental Fig. S1**). Our device consists of a uniform 1.7 µm thick layer of NV centers in a 4 × 4 mm$^2$ diamond illuminated with a laser from an oblique angle (**Fig. 1 (b)**). The resulting fluorescence containing information on the magnetic field is imaged onto a CCD, providing a two-dimensional (2D) image of the field. The diamond is placed directly on top of the device under test (DUT) to minimize standoff distance. Advantages of this technique over other magnetometers include simultaneous high spatial resolution (~1 µm), wide field-of-view (few mm), and vector magnetic field imaging under ambient conditions. Our previous QDM studies demonstrated mapping of the functional activity of both wirebonded and flip chip integrated circuits via frontside and backside magnetic field imaging, respectively [15,16].

Here, we show the QDM's potential as a failure analysis tool for modern, complex ICs and the next-generation of 3D microelectronics. We measure the magnetic fields produced by currents in a state-of-the-art flip chip IC as well as 3D current paths in a 3D IC analog built out of a thin, multi-layer PCB. For both devices, we measure all three vector components of the magnetic field over a 4 × 4 mm$^2$ field of view with ~1 µm lateral spatial resolution. The increased data capture enables the identification of vertical current paths and improved 2D localization. For the flip chip, our QDM detects the magnetic field produced by the C4 bumps, leading to an *in-situ* localization method circumventing the need for optical and magnetic image alignment. With this localization, we are able to correlate the current flow with provided CAD images. For the PCB, we detect current traveling along a 200-µm-long vertical via as well as isolating a plane of current density from three different active layers through the linearity of Maxwell's equations. In addition, we verify the separation between two conducting layers from the detected magnetic field. We place these results in the context of failure analysis and discuss future improvements to the tool that will enable increased throughput and detection of open faults.

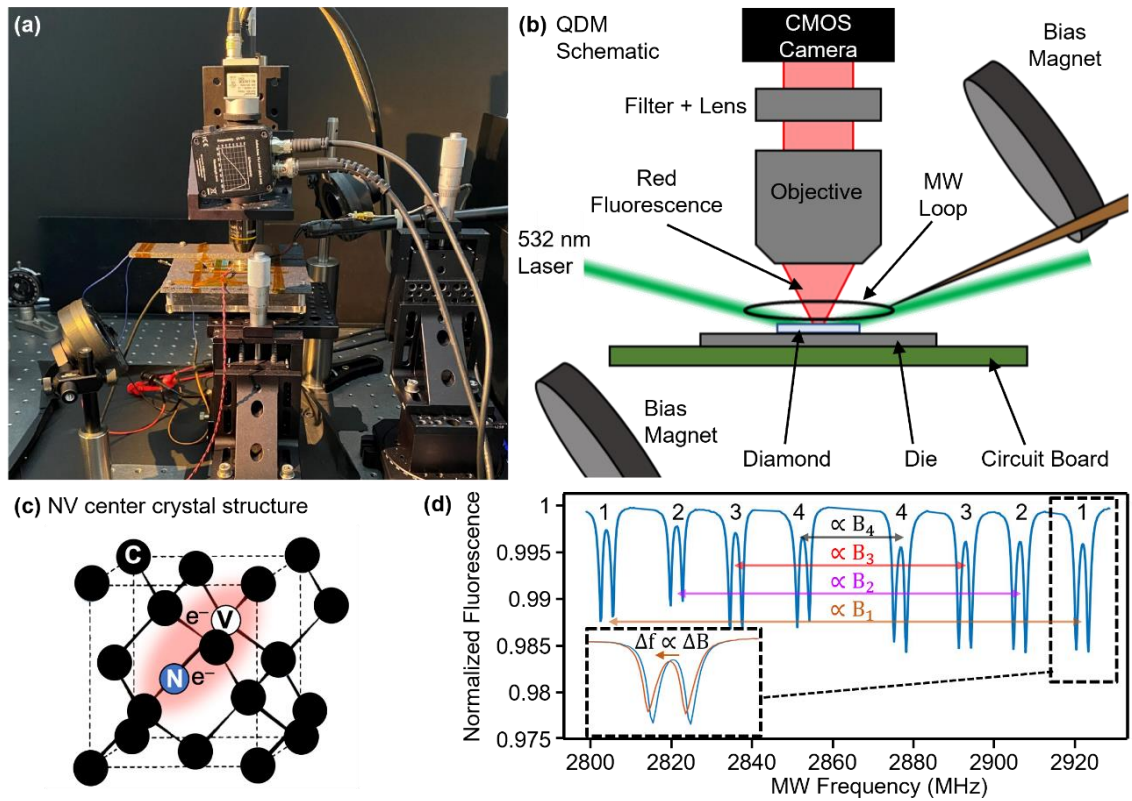

**Figure 1. (a)** Photo and **(b)** schematic of the quantum diamond microscope (QDM). Key components include a diamond with a thin surface layer of nitrogen-vacancy (NV) centers, which is placed directly on top of the device under test (DUT); a 532 nm excitation laser; a microwave (MW) loop; bias magnets; and imaging optics consisting of an objective, filter, lens, and CMOS camera. **(c)** The crystal structure of an NV center, which consists of a substitutional nitrogen atom (N) and a neighboring lattice vacancy (V) in the diamond's carbon lattice. There are four possible orientations of the NV axis in the diamond lattice, which gives rise to the QDM's ability to simultaneously obtain full vector magnetic field information in a single measurement as the field projects differently onto each axis. **(d)** An example of the QDM's optically detected magnetic resonance (ODMR) spectroscopy measurement technique, which involves exciting the NV-diamond with a 532 laser and monitoring the resultant red fluorescence as a function of MW frequency applied using the MW loop. Dips in NV fluorescence correspond to MW frequencies resonant with the NV electron spin transition energies, which split due to the NV Zeeman interaction with the external magnetic field (see NV energy diagram in **Supplemental Fig. S1** for more detail). The frequency separation between peak pairs labeled 1-4 is proportional to the component of the vector magnetic field parallel to one of the four NV axes (labeled $B_1$ - $B_4$). Bias magnets initially separate the spin transition energies, and the QDM then images spatial variations in the magnetic field $\Delta B$ from the DUT by detecting small ODMR frequency shifts $\Delta f$ as shown in the inset.

## Experimental Methods

As shown in **Figs. 1(a)** and **1(b)**, the QDM consists of a pair of 2" SmCo bias magnets, a 7 mm diameter microwave (MW) loop antenna delivering 1 W of MW power to the NVs, a microscope objective (we use both a 4× magnification, 0.1 NA objective and a 20× magnification, 0.35 NA objective) to collect NV fluorescence over the diamond field-of-view, a CMOS camera to measure NV fluorescence, a 532 nm laser delivering 1.5 W continuous wave (CW) illumination over a 4 × 4 mm² area, and a 4×4×0.5 mm³ diamond chip, with NVs embedded within a 1.7 μm isotopically pure layer of [$^{12}$C]~99.995 %, [$^{15}$N]~17 ppm, and [NV$^-$]~2 ppm.

The QDM measures magnetic fields with a technique known as optically detected magnetic resonance (ODMR) spectroscopy (an example ODMR spectrum is shown in **Fig. 1(d)**). A QDM measurement protocol involves placement of the diamond chip directly on top of the DUT (see **Fig. 1(b)**) to minimize sensor standoff distance and maximize spatial resolution. The 532 nm laser, modified with beam shaping optics into a top-hat profile, illuminates the entire diamond with equal laser power at a shallow angle to the sample surface. The NV centers in the diamond surface layer fluoresce with red photons (637 - 800 nm) as they are initialized into an optically bright electron spin state. NV fluorescence is collected with the microscope objective, filtered with a 633 nm longpass filter, and imaged onto the CMOS sensor. Fluorescence intensity is monitored across the entire diamond simultaneously as a function of the applied MW frequency provided to the loop antenna. When the MW frequency is resonant with an NV electron spin transition, there is a decrease in fluorescence intensity (dips in ODMR spectra shown in **Fig. 1(d)**) as the excited NV electronic spin states relax through an alternative decay pathway (see **Supplemental Fig. S1** for more detail). The ODMR spectra are fit with inverted Lorentzians to determine the frequencies where dips in fluorescence intensity occur. This is then converted to magnetic field strength, as frequency separation in the ODMR spectra is proportional to the magnetic field parallel to the NV axes at a given pixel, given by the quantum physics of the NV spin energy levels. There are four possible orientations of NV axes in the diamond lattice — with ODMR spectral shifts for each NV orientation class dependent on the projection of the ambient magnetic field along the respective NV axis — which enables simultaneous measurement of the x, y, and z components of the magnetic field (referred to as $B_x$, $B_y$, and $B_z$). Projection of a vector magnetic field can be different along each of the four NV axes, as demonstrated by the differing frequency separations between intensity dips labeled with number pairs 1-4 in the ODMR spectrum of **Fig. 1(d)**. Magnetic field distributions from the DUT are calculated by fitting the ODMR spectra at each pixel and subtracting the magnetic field of the bias magnets. For a detailed flowchart on the QDM measurement protocol, see **Supplemental Fig. S2**.

To explore the QDM's ability to resolve current in complex structures, we investigate two samples in this study. We first explore an NVIDIA GA106 graphics processing unit (GPU, **Fig. 2(a)**) to test lateral spatial resolution and the ability to detect current traces in a modern and complex flip chip IC. The GPU is 276 mm$^2$ with 13,250 million transistors fabricated using an 8 nm production process at Samsung. The die is thinned to 5 µm remaining silicon thickness (RST) to reduce the standoff distance of the diamond from the current sources. To drive current through the device, the chip is biased through the JTAG pins on the PCB to which it is mounted. To recreate aspects of a 3D IC's architecture, we next use a custom four-layer PCB (Advanced Circuits) for studying the QDM's ability to image magnetic fields from a 3D structure. The board is ~5.8×5.8 cm$^2$ and has four layers of current traces with varying layouts that are interconnected in some cases with vias. See **Fig. 4(a)** for dimensions of the layers. Magnetic inverse calculations to determine the underlying current distributions are performed in MATLAB using the Fourier Filter formalism [17].

## Results and Analysis

We first present the QDM's ability to image vector magnetic fields of a modern IC with a wide field-of-view and high spatial resolution under ambient conditions. The wide field imaging capability of the QDM is in stark contrast to scanning measurements, such as those done with SQUID and GMR magnetometers, thus allowing the QDM to do simultaneous imaging of currents over a several millimeter field-of-view at ~kHz frame rates [18]. This capability is first demonstrated in this study using an NVIDIA GA106 GPU shown in **Fig. 2(a)**.

Spatial resolution of magnetometers decreases with increasing standoff distance of the sensor from the source since magnetic fields decrease with the inverse square law. Thus, the NVIDIA GA106 GPU's die is thinned to 5 µm RST to improve spatial resolution. We note that in a previous study we demonstrated QDM magnetic field imaging of active ring oscillators (ROs) in a field-programmable gate array (FPGA) at a much larger standoff distance of ~500 µm [15]. While spatial resolution declined at larger standoff distances, machine learning techniques were employed to interpret the images and were successfully used to localize and identify the number of ROs programmed to be active on the FPGA.

JTAG pins on the NVIDIA chip are biased to drive current through the die for magnetic field imaging with the QDM. Biasing of the Test Rest pin allows current to flow through the path labeled TRST_N in the CAD image of **Fig. 2(b)**, while biasing the Test Data Out pin results in current flow through the trace labeled TDO. Both current paths are on the edge of the die (location marked by the red dot in **Fig. 2(a)**), so we place the diamond on top of this location and slightly cantilevered off the die, as demonstrated in the schematic in **Fig. 2(a),** to entirely image their magnetic fields. We note that the diamond NV layer is 1.7 µm thick, and since the diamond sits directly on top of the DUT, the average sensing standoff distance (each individual NV center) is roughly half this thickness (0.85 µm) from the top surface of the DUT. This metric is greatly improved over SQUID magnetometers, which have typical standoff distances of ~100s μm [9]. While SQUID magnetometers currently have better sensitivity than that of NV ensembles (~pT/√Hz for SQUID vs. ~10 pT/√Hz for NVs [19]), the application of a SQUID is challenged by the minimum standoff distance set by the cryogenic housing of the sensor head. GMR sensors, on the other hand, can have a standoff distance of ~1 μm from the sample, but their sensitivity (~1 nT/√Hz) is worse than that of both NV-diamond and SQUID magnetometers [9].

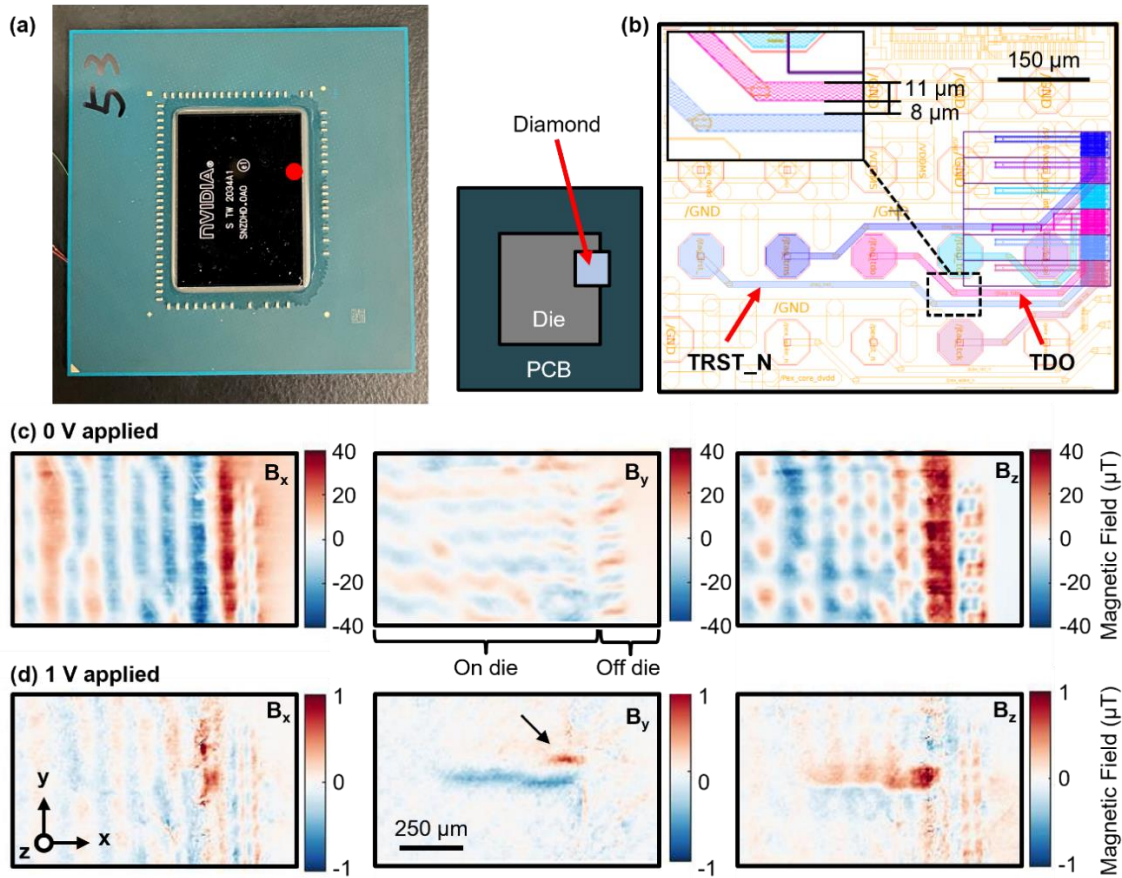

**Figure 2. (a)** Photo of the NVIDIA GA106 GPU used in this study. The image was taken before thinning the die to 5 μm remaining silicon thickness. The red dot corresponds to the location where QDM measurements were taken. The 4× 4 mm$^2$ diamond is placed directly over the red dot with the diamond slightly cantilevered off the side of the die as shown in the schematic to the right of the image. **(b)** A CAD drawing of the JTAG traces inside the silicon die from which current is measured. To drive current through these paths, the JTAG Test Reset and Test Data Out pins are biased, which allows current to flow through the traces labeled TRST_N and TDO, respectively. The inset is from the dashed box, which illustrates that the TRST_N and TDO current paths are 11 μm wide and have minimum separation of 8 μm. Panels **(c)** and **(d)** show the B$_x$, B$_y$, and B$_z$ QDM magnetic field images taken in the same location on the chip when it is either unbiased (0 V applied) or when the TRST_N path is biased with 1 V, respectively. The TDO path is biased in later experiments. Note the large magnetic field gradients that exist on the left side of the figures in panel **(c)** are due to magnetic material present in the flip chip's C4 bumps. These large gradients are not present on the right side of the magnetic field images since the diamond is cantilevered off the edge of the die, as demonstrated by "On die" and "Off die" labels in the center figure of panel **(c)**. The TRST_N current traces are resolved all the way to the edge of the die in the B$_y$ and B$_z$ magnetic field images of panel **(d)** after subtracting the unbiased magnetic field images of panel **(c)** from magnetic field images when the chip is biased with 1 V. The black arrow highlights magnetic field signal from current in the JTAG pad. The scale bar in panel **(b)** is 150 μm and the scale bar for panels **(c)** and **(d)** is 250 μm.

Before driving current through the NVIDIA chip, we use the QDM to image the vector magnetic field of the sample's die over the JTAG current paths highlighted in **Fig. 2(b)**. The B$_x$, B$_y$, and B$_z$ components of the magnetic field can be seen in **Fig. 2(c)**. Large magnetic field gradients due to the flip chip C4 bumps dominate the left side of the images where the diamond rests on top of the die. Magnetic field contributions of C4 bumps in QDM images were investigated in detail in our previous study [16]. Energy-dispersive X-ray spectroscopy measurements indicated the presence of elemental Si, Sn, Al, Ni, Cu, and Ti in and around the C4 bumps. Follow-up finite element analysis modeling showed Ni in the under-bump metallization is magnetized in the presence of the QDM's bias magnet field, which manifests as large magnetic field gradients. While these effects can make it difficult to resolve very small currents in the die, they can be subtracted from the magnetic field images during data analysis; and also can be used for image registration to determine the location of the diamond on the die relative to current distributions embedded in the device, which is not a trivial task (see **Supplemental Fig. S3**).

Next, we bias the NVIDIA GPU to drive current through the TRST_N trace shown in **Fig. 2(b)**. This allows us to investigate the ability of the QDM to resolve a magnetic field signal of a current path amongst the large background gradients from the C4 bumps commonly present in modern ICs. The current path travels from the PCB and then up into the die where it passes through a weak pull-down resistor. The $B_x$, $B_y$, and $B_z$ magnetic field images of the current trace in the die when biased with 1 V are shown in **Fig. 2(d)**. For an overlay of the $B_z$ magnetic field image on the CAD drawing, see **Supplemental Figure S4(a)**. The trace has a width of ~11 µm. We remove signals from C4 bumps by subtracting a magnetic field image of the unbiased chip from the that of the biased chip. This approach has improved over our previous work where we found it challenging to remove these features [16], which we now accomplish through improvements in data analysis and instrumentation. For initial experimental exploration, a thermistor and PID controller are used to keep the sample at a constant temperature to help isolate the signals of interest. Without temperature stabilization, biasing of the current paths results in local Joule heating that scrambles the magnetic domains in the Ni under bump metallization and results in a modified background magnetic field gradient that can be challenging to ameliorate. For the data presented in **Figs. 2** and **3,** the magnetic contributions of the C4 bumps were removed through the use of high SNR no-current reference measurements to simplify the experimental protocol. The magnetic field from the current is resolved in the $B_y$ and $B_z$ images with a magnitude of ~1 µT and it extends all the way to the edge of the die. Current from the JTAG pad is also resolved in the $B_y$ image (black arrow in **Fig. 2(d)**). Since the current path is parallel to the x direction (see compass in **Fig. 2**), there is no $B_x$ component for this current as expected from the right-hand rule for magnetic fields. Other signals appear in the $B_x$ magnetic field image, which are likely due to local Joule heating from the current trace causing a change in the magnetic properties of the nearby C4 bumps. We note that while the ~1 µT field in the $B_y$ and $B_z$ images is large in comparison to the ~10 pT fields detectable by NV ensembles, we used the QDM in a previous study to detect ~100 pT to ~1 nT fields from ROs on an FPGA [15].

To investigate QDM spatial resolution for nearby, in-place current traces, we drive current through both the TRST_N and TDO current paths in **Fig. 3**. The separation between the TRST_N and the TDO paths is 8 µm, as shown in the inset of **Fig. 2(b)**. The ability to resolve nearby traces is effective for detecting shorts and leakages in a faulty device. The vector magnetic field components of the TRST_N trace when reverse biased with -0.15 V is shown in **Fig. 3(a)**. The polarity of the magnetic field is reversed in the $B_y$ and $B_z$ magnetic field images in comparison to **Fig. 2(d)**. To isolate the magnetic field signal from current flowing through the TDO trace, we bias this path with -0.15 V. The vector magnetic field components of the TDO trace are shown in **Fig. 3(b)**. For an overlay of the $B_z$ magnetic field image of the TDO trace on the CAD drawing, see **Supplemental Figure S4(b)**. This trace, which has a width of 11 µm, is resolved in the $B_y$ and $B_z$ images and a small $B_x$ component is detected as well where the path is angled at 45° to the x axis (black arrow). Both the TRST_N and TDO paths are simultaneously biased with -0.15 V in **Fig. 3(c)**, and we find that the magnetic field emanations of the TRST_N path overwhelm the signal from the TDO path. However, as shown in **Fig. 3(d)**, we find that it is possible to differentiate between the two current traces if we bias the TRST_N path with 1.0 V and reverse bias the TDO path with -0.15 V. This ability to resolve the magnetic field signatures of adjacent traces on the micron scale shows the potential for the QDM to detect shorts and leakages in a faulty device.

Due to the increasing demand for failure analysis techniques to detect faults in 3D structures, we designed and printed a custom multi-layer PCB for QDM magnetic field imaging of 3D current distributions (**Fig. 4(a)**). As shown in the schematic in the right side of **Fig. 4(a)**, the total board thickness is 290 µm and it has 4 layers of copper traces. The copper traces in layers 1 and 4 are 13 µm tall and those in layers 2 and 3 are 14 µm tall. In the xy-plane, all layers are 76 µm wide and have a minimum separation of 76 µm. Layers 1 and 2 and layers 3 and 4 have the same separation of 49 µm, while the larger separation between layers 2 and 3 is 102 µm due to the presence of a stabilizing core. The layout of current paths in individual layers varies, and in some cases is interconnected with vias that have a diameter of 152 µm. An 18 µm thick protective soldermask covers the top and bottom of the board. For electrical connection to the copper traces, we soldered a terminal block with push-in tension clamps to the PCB. The 4×4 mm² white squares printed on the PCB that can be seen in the left-hand image of **Fig. 4(a)** indicate locations below which we have designed structures to test various aspects of 3D microelectronics. The diamond is placed directly on top of these regions for mapping of the current distributions below via QDM magnetic imaging.

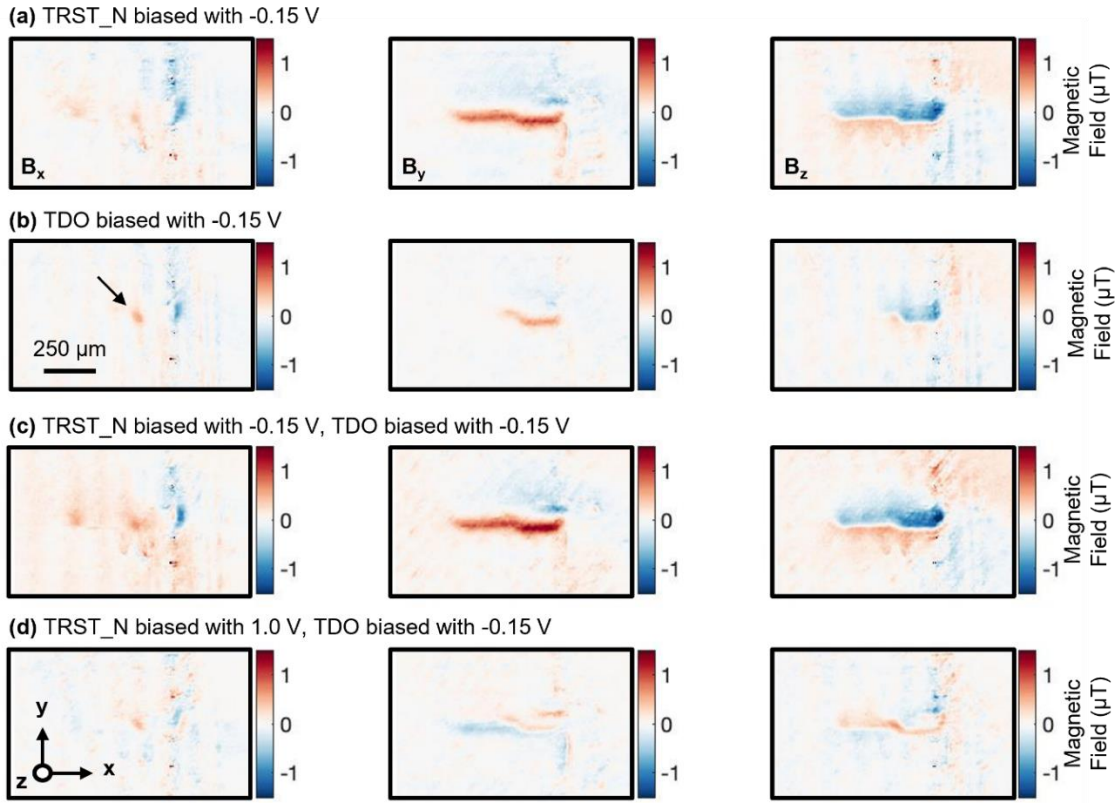

**Figure 3.** QDM vector magnetic field images ($B_x$, $B_y$, $B_z$) of the NVIDIA GA106 GPU when different combinations of the JTAG Test Reset and Test Data Out pins are biased, which results in current flow through the traces labeled TRST_N and TDO in the CAD image in **Fig. 2(b)**. **(a)** TRST_N is reverse biased with -0.15 V, showing that the polarity of the magnetic field is reversed in comparison to **Fig 2(d)**. **(b)** TDO is reverse biased with -0.15 V to demonstrate that this current path can be resolved in the QDM magnetic field images. The black arrow in the $B_x$ image highlights magnetic field signal from the trace when it is angled at 45° to the x direction. **(c)** TRST_N and TDO are simultaneously biased with -0.15 V. The magnetic field from the TRST_N current trace overwhelms the signal from the TDO trace. **(d)** TRST_N is biased with 1.0 V and TDO is biased with -0.15 V to demonstrate that the individual current paths can be resolved, as shown in the $B_y$ magnetic field image. The scale bar is 250 µm and applies to all panels.

Simultaneous measurement of the vector components of magnetic fields can be used to detect vertically oriented current paths, such as vias, which we demonstrate in **Fig. 4(b)**. The leftmost panel shows the schematic for structures located in the red dashed box labeled "b" in **Fig. 4(a)**. There are three side-by-side traces in which current can pass through vias from the top layer 1 (blue traces) to either layer 2 (orange trace), layer 3 (red trace), or layer 4 (magenta trace). The yellow circles are the vias. Current paths oriented in the vertical z direction have only $B_x$ or $B_y$ magnetic field components. Detection of such vertical currents are challenging for other magnetometers, such as the SQUID and GMR sensors, that only measure the out-of-plane component of the magnetic field ($B_z$). In **Fig. 4(b)**, current is driven only through the center trace shown in the schematic. Consider first the QDM $B_z$ magnetic field image in the rightmost figure. Due to the right-hand rule for magnetic fields, the $B_z$ component should be zero directly over a current path so that traces can be followed by looking for where the magnetic field images read zero. As the current passes from right to left in layer 4, the magnetic field of the current is detected but it is small and is significantly blurred. When the current passes through the via from layer 4 to layer 1, the magnetic field goes to zero, and when the current flows from the via to the left, the spatial resolution of the current path is improved and the intensity of the magnetic field is much higher. For techniques that only measure this out-of-plane component of the magnetic field, vias are detected by the discontinuity in current, which can be difficult to interpret as the magnetic fields of an arbitrarily large number of current sources can sum to zero. However, the QDM's ability to simultaneously measure all three vectorial components of a magnetic field ($B_x$, $B_y$, $B_z$) provides more information and allows for direct detection of current flowing through vias, as shown in the two middle panels of **Fig. 4(b)**. In the case of the $B_y$ image, the via is revealed by a dipole-like feature (zoomed in inset). The dipole-like signal is overwhelmed in the $B_x$ image due to the component of the current path just above the via that displays a significant $B_x$ magnetic field, but a $B_x$ field can still be seen around the outside edge of the via.

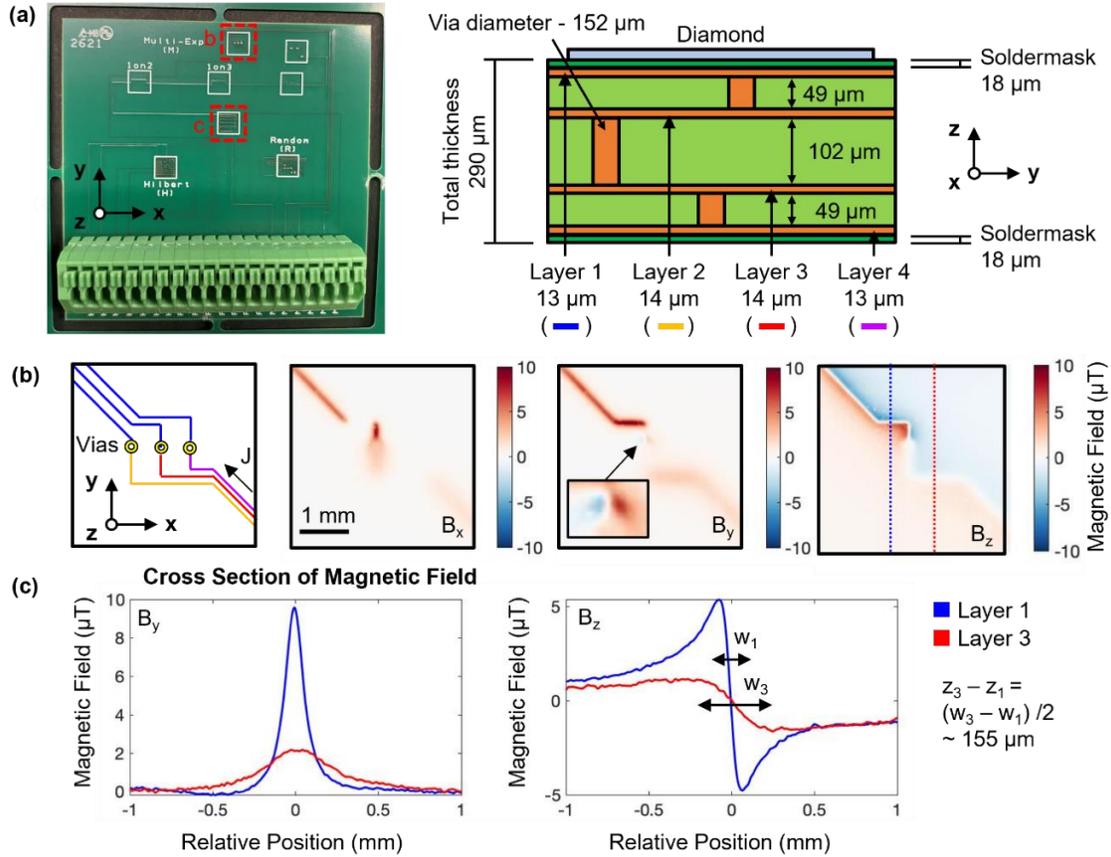

**Figure 4.** (a) For benchmarking the QDM's ability to image three-dimensional (3D) current paths, we designed and printed a custom 4-layer printed circuit board (PCB) with various experiments in locations indicated by white squares printed on the board's silkscreen. A schematic of the board as seen from the side is shown on the right-hand side of panel (a). The layers are labeled 1, 2, 3, and 4 (colored bars correspond to the schematic in panel (b)) and have different distributions of copper traces in each, which in some cases are interconnected with copper vias. A protective soldermask encapsulates the bottom and the top of the board. For magnetic field sensing of the current paths, a diamond with a thin layer of NV centers is placed directly on top of the board. (b) A schematic of 3D current traces on the PCB below the red dashed box labeled "b" in the left photo of panel (a). Each current trace is 76 µm wide and has a minimum separation of 76 µm. Current passes through the vias from layer 1 (blue traces) to either layer 2 (orange traces), layer 3 (red traces), or layer 4 (magenta traces). We drive current through the center path by biasing with 100 mV. The QDM $B_x$, $B_y$, and $B_z$ magnetic field images to the right show blurring of the magnetic field for the current traces in layer 3 relative to layer 1. The QDM's ability to simultaneously measure $B_x$, $B_y$, and $B_z$ magnetic field components allows it to detect current contributions from the vias. No significant magnetic field is detected from current in the via in the $B_z$ image, whereas a magnetic field is detected around the via in the $B_x$ image and the field from the via appears as a dipole in the $B_y$ image (see zoomed-in inset). The scale bar in panel (b) is 1 mm. (c) We plot line-slices of the magnetic field image for the y-and z-component of the magnetic field to illustrate the depth-localization capabilities of the QDM. The separation of the maxima and minima of the z-component of the magnetic field can be used to approximate the depth of wire-like current sources [20]. Current flowing in layer 1 (blue traces) is steeper and narrower for both the y and z components of the magnetic field compared to layer 3 (red traces). The separation of layer 1 and layer 3 is determined to be $(w_3-w_1)/2 \sim 155$ µm, which is consistent with the separation of layer 1 and layer 3 from the schematic in (a).

For wire-like current sources, the z-depths of the current sources can be determined through the spread of the maxima and minima in the $B_z$ component of the magnetic field [18]. This established technique from Scanning SQUID Microscopy is utilized with the QDM on the 3D PCB, where we have full knowledge of the location and path of the relevant traces. The $B_z$ via data in **Fig. 4(b)** illustrates the difference in the amplitude and spread of the magnetic field when comparing the current in the top layer and the third layer. Line slices indicated by the dashed-lines in **Fig. 4(b)** are plotted in **Fig. 4(c)** to illustrate the depth dependence of the magnetic field profile. The positions of the magnetic field cross sections are offset to allow for direction comparison of the distribution. The location of the maxima and minima are found through peak-finding and the difference in the widths is found to be ~310 µm, indicating a difference in depths of ~155 µm. This number is consistent with the design specifications of the PCB in **Fig. 4(a).** Further work will also use the broadening of the $B_x$ and $B_y$ components of the magnetic field to give two different estimates of the depth, exploiting the magnetic vector imaging capabilities of the QDM.

**Figure 5(a)** shows a schematic of the 3D current paths in the region of the PCB outlined with the red dashed box labeled "c" in the photo of **Fig. 4(a)**. There are three layers of current traces parallel to each other: the blue traces are in the topmost layer 1, the orange traces are 49 µm below in layer 2, and the red traces are 102 µm deeper into the PCB in layer 3, as shown in **Fig 4(a)**. Note that the minimum separation between traces in the xy-plane in any of these layers is 76 µm. We designed these traces so that current in each layer can be activated independently. **Figures 4(b-d)** show the QDM $B_z$ magnetic field images when only layer 1, layer 2, or layer 3 are biased with 100 mV. Due to the inverse square law, signal from layers embedded deeper in the structure show a weaker field and are blurred. **Figure 5(e)** shows a $B_z$ magnetic field image of the same circuit when layers 1, 2, and 3 are all biased with 100 mV simultaneously.

The industry's trend towards adoption of 3D microelectronics packaging points to the need for failure analysis engineers to extract z-depth information of faults in complex 3D structures. Magnetic field imaging is effective for detecting current distributions in 3D circuits since magnetic field lines penetrate materials that other techniques such as optical probing cannot. However, interpretation of magnetic field images by non-experts is challenging and extracting quantitative data necessary to obtain z depth information conveys the need for solving the 3D magnetic inverse problem. The 3D magnetic inverse problem refers to inversion of the Biot-Savart Law for a magnetic field **B**, given by the equation:

$$\mathbf{B}(\mathbf{r},t) = \frac{\mu_0}{4\pi} \int d^3 r' \frac{\mathbf{J}(\mathbf{r}',t) \times (\mathbf{r}-\mathbf{r}')}{|\mathbf{r}-\mathbf{r}'|^3},$$

to solve for current density **J**. Here, **r** is the point of observation, t is time, and $\mathbf{r}'$ is a point on the current source. Note that this is known to be an ill-posed problem so that a unique solution does not exist: an arbitrarily large number of current configurations can sum together to produce the same magnetic field. Different techniques have been developed in attempts to constrain and solve such ill-posed inverse problems, including Tikhonov regularization [21], Fourier Filters [17,22], estimation theory [23], probabilistic multi-source reconstructions [22], least square fitting [25,26], Bayesian methods [27], genetic algorithms [28], and direct mapping and fitting in low dimensional systems [29–31]. Particularly in the case of failure analysis, previous approaches have tried to side-step this obstacle by assuming specific geometries and utilizing prior knowledge of the DUT to map 3D current distributions [32–34].

Importantly, a unique solution to the magnetic inverse problem exists if the problem is constrained to two dimensions. The 2D magnetic inverse problem has been shown to be straightforward to solve, as has been demonstrated for conversion of SQUID magnetic field images to 2D current density maps in Refs. [9] and [12]. Since inversion of the Biot-Savart Law includes a complicated integral, the standard approach is to do the operations in Fourier space, which reduces the problem to simple multiplication and division [17].

In the present study, we make initial steps to generate current density maps for sources in 3D structures. We utilize the linearity of Maxwell's equations to subtract the magnetic field images of layers 1 and 3 in **Figs. 5(b)** and **5(d)** from the magnetic field image of the 3D circuit when all layers are active in **Fig. 5(e)**. The resultant magnetic field image after subtraction is shown in **Fig. 5(f)**, which reproduces the magnetic field image of **Fig. 5(c)** where only layer 2 is active. Noise and blurring are slightly amplified and remnants of the traces from layer 1 are present. Nonetheless, we show this approach is sufficient to generate a current density map via the 2D magnetic inverse for layer 2 embedded 67 µm in the structure (soldermask plus PCB layering), which is plotted in **Fig. 5(g)**. Note that this result reproduces the orange traces in the schematic of **Fig. 5(a)**.

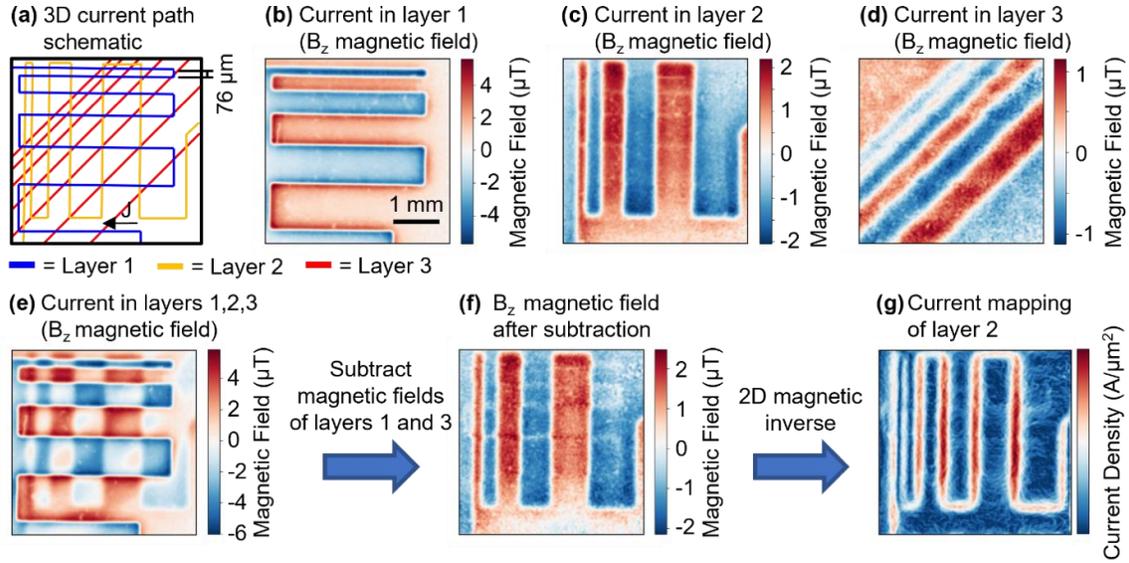

**Figure 5.** (a) Schematic of 3D current paths in the custom, multi-layer PCB designed for this work. These structures are located in the red dashed box labeled "c" in the photo of **Fig. 4(a)**. The blue traces are in the topmost layer 1, the orange traces are 49 µm below in layer 2, and the red traces are in layer 3, which is 102 µm below layer 2. See the schematic of **Fig. 4(a)** for more details of the PCB. QDM $B_z$ magnetic field images when current flows only through (b) layer 1, (c) layer 2, or (d) layer 3. In all cases, the current paths are biased with 100 mV. (e) $B_z$ magnetic field image when layers 1, 2, and 3 are biased with 100 mV. (f) Resultant $B_z$ magnetic field image after subtracting the magnetic field images of layers 1 and 3, corresponding to panels (b) and (d), respectively, from the image of panel (e) where current is active in all layers. The final image represents the magnetic contribution from layer 2 only, as seen when comparing panel (f) to panel (c). (g) Current density map of layer 2 generated via the two-dimensional (2D) magnetic inverse problem using the magnetic field image of panel (f). This demonstrates that the linearity of Maxwell's Equations can be used to subtract magnetic field contributions from individual layers to map current distributions embedded in a 3D structure. The scale bar is 1 mm and applies to all panels.

The ability to subtract magnetic field images from each other and then solve the 2D inverse problem suggests a protocol for providing z depth information for 3D current distributions. Neural networks are proving to be advantageous for analyzing and interpreting multi-dimensional datasets and even ill-posed problems [35–37]. This can be a useful tool for solving inverse problems, especially since the forward problem is well described by the Biot-Savart Law and is made more computationally efficient with the Fourier Filter formalism [17]. The accuracy of the neural network output depends on the quality and amount of training data, which can be supplemented with prior information for ICs alongside classes of current sources, noise levels, and standoff distances to optimize performance. The idea is to use neural networks to determine magnetic field contributions from individual layers of current sources and subtract them from the magnetic field image of the 3D structure in an iterative process. The neural network can solve the 2D magnetic inverse problem to find the current distribution in a single layer, and then the forward problem (determining the magnetic field from the current density) can be calculated to determine the magnetic field for that layer, which can then be subtracted from the magnetic field image of the 3D structure. In this way, current distributions from individual layers can be isolated, which when coupled with prior knowledge of the circuit can provide z depth information for current paths and thus the ability to map current distributions in embedded layers. We recently made encouraging initial progress using neural networks for identifying current distributions in two dimensions [18], and our research into using this approach for 3D structures is ongoing.

## Conclusion and Outlook

In this work, we report magnetic imaging of 2D current paths in a flip chip IC and 3D current paths in a multi-layer PCB using a quantum diamond microscope (QDM). The results demonstrate the QDM's ability to obtain simultaneous wide field-of-view, high spatial resolution, vector magnetic field imaging under ambient conditions. Magnetic field imaging of flip chips can be complicated by the large field gradients produced by magnetic material in the C4 bumps, but they can also be used for image registration; also our analysis techniques allow removal of their contributions to reveal signal from current paths in the die. We explore forward and reverse biasing of different JTAG pins in the flip chip to drive current through nearby traces in the die, and show that adjacent current traces with a width of 11 µm and a separation of 8 µm can be individually resolved. We also design and print a 4-layer PCB to benchmark the QDM's magnetic field imaging capabilities against 3D current distributions. Experiments can be designed to allow for separate biasing of current paths in individual layers and to explore the magnetic signatures of vias. The vertical orientation of vias results in no z component of the magnetic field, as we verify in the QDM's $B_z$ magnetic field images. This makes it challenging for other magnetometers to detect current in vias since they typically measure only the z component of the magnetic field. Here, however, we demonstrate the QDM's ability to detect vias in $B_x$ and $B_y$ magnetic field measurements. The 3D current distributions in the PCB also allow demonstration of the effects of standoff distance on spatial resolution and signal strength detected as well as confirm the ability to distinguish relative vertical distances between different conducting layers.

One of our goals with the QDM is to provide z depth information for 3D current distributions. Such a capability would allow for identifying and mapping current sources in complex 3D structures and providing important quantitative information to failure analysis engineers necessary for next-generation ICs. In our initial steps towards this ability, we image magnetic fields from current paths in a 3-layer structure and reconstruct current traces from the middle layer embedded 67 µm in the sample. We utilize the linearity of Maxwell's equations to subtract magnetic field images of the top and bottom layer from that of the total structure; and then solve the 2D magnetic inversion problem to map the current density of the middle layer. These results point to future capabilities of using neural networks integrated with prior information of ICs and coupled with solution of the 2D magnetic inversion to isolate magnetic field contributions and map current distributions from separate layers of a 3D IC. We previously engaged in initial work into such a machine learning approach [18], and related research is ongoing.

The sensitivity of NV-diamond magnetic sensors is improving every year as the technology matures [16,36]. Continued technology improvements will also further advance the capabilities of magnetic field mapping with wide-field QDM measurements. Therefore, as we further develop the QDM's capabilities and the component technologies continue to improve, we believe this technique will become a valuable tool to the IC failure analysis community for measurement of shorts, leakages, and potentially open failures.

## Acknowledgements

We thank NVIDIA for providing test devices and design information.

This work was funded by the MITRE Innovation Program. M.J.T. and R.L.W. acknowledge support from the Quantum Technology Center (QTC) at the University of Maryland. NV-diamond sensitivity optimization pertinent to this work was partially supported by the DARPA DRINQS program (Grant No. D18AC00033).

# Supplemental Material

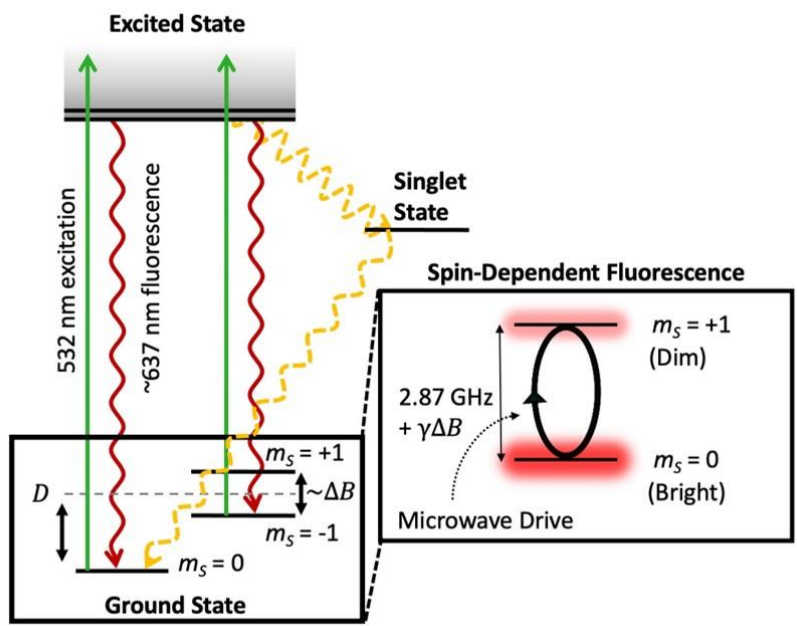

**Figure S1.** Nitrogen-vacancy (NV) center energy diagram. The ground state of an NV electron is a spin triplet with electron spin $m_s = 0$ or $m_s = \pm 1$. In the absence of a magnetic field, the $m_s = \pm 1$ sublevels are energy degenerate and are separated from the $m_s = 0$ level by the zero-field splitting term $D$. This term is temperature dependent and has a value of ~2.87 GHz at room temperature. Excitation of electrons from the ground state to the excited state with a 532 nm excitation laser results in red fluorescence (637 - 800 nm emission) as the electrons relax back down to the ground state. Spins in the excited state with $m_s = 0$ preferentially decay directly to the ground state through the emission of red fluorescence and without change of the spin state. However, electrons with $m_s = \pm 1$ may decay through the singlet state back down to the ground state (leading to decreased red fluorescence) and with change of the spin state to $m_s = 0$. In the presence of a magnetic field, the energy degeneracy of the $m_s = \pm 1$ spin states is lifted through the magnetic Zeeman interaction, labeled $\Delta B$ in the figure above.

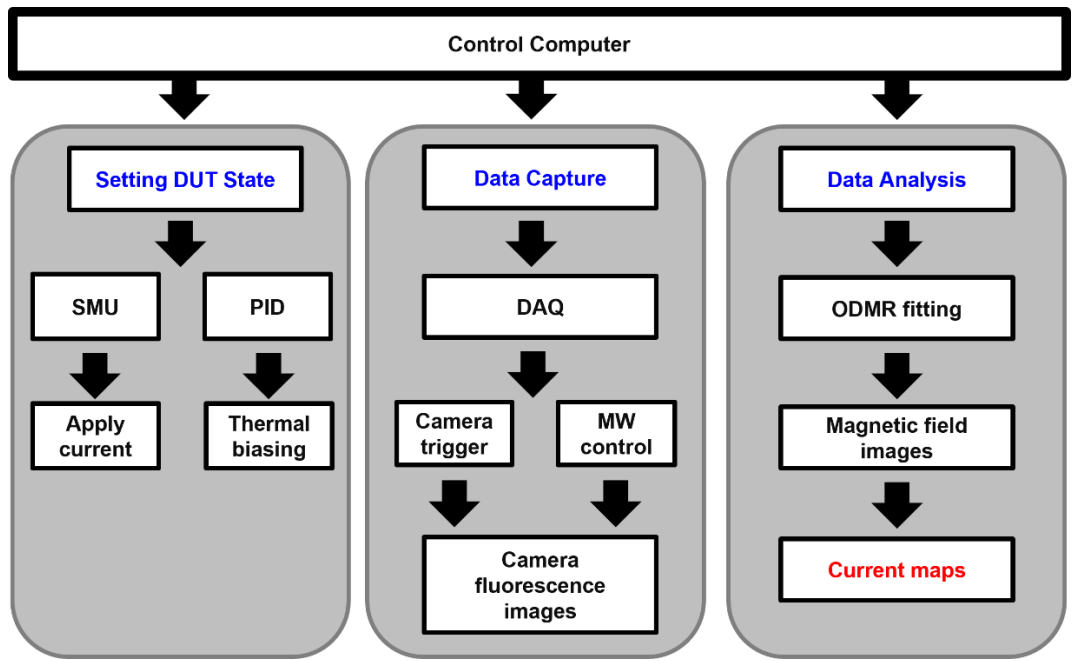

**Figure S2.** Flowchart of the QDM magnetic imaging experiment protocol. A central control computer interfaces with the device under test (DUT) to either source voltage/current for biasing the chip or for control of an external PID loop for thermal biasing as needed. The same central computer controls the individual components of the QDM including the DAQ, camera, and MW parameters. Finally, the computer analyzes the data to extract magnetic field maps from shifts in the ODMR resonances. Current maps are made by inverting the magnetic field images.

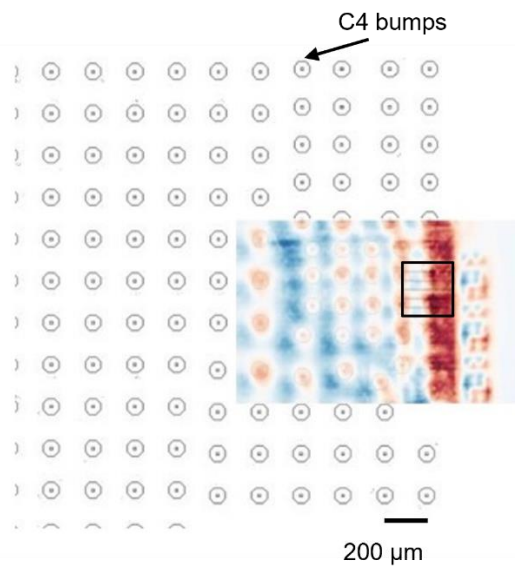

**Figure S3.** Overlay of a QDM $B_z$ magnetic field image on top of a CAD drawing of the NVIDIA GA106 GPU C4 bumps to demonstrate the ability to use the signal from these features for magnetic field image registration. The rectangles in the CAD drawing are the JTAG pads seen in **Fig. 2(b)**. The scale bar is 200 µm.

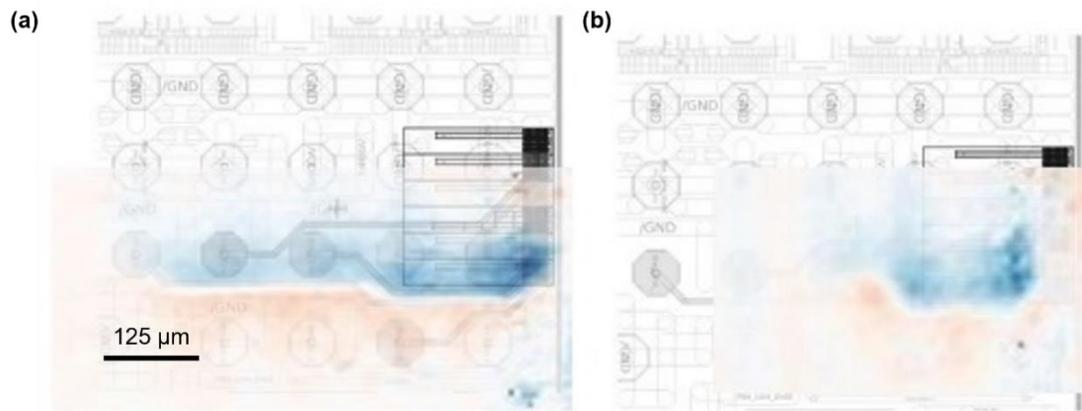

**Figure S4.** Overlay of QDM $B_z$ magnetic field images on top of the CAD drawing of the NVIDIA GA106 GPU current traces when the **(a)** TRST_N current path is biased with -0.15 V and the **(b)** TDO current path is biased with -0.15 V. The scale bar is 125 μm.